\documentstyle[epsf,amssymb]{elsart}

\def\bfnabla{\mbox{\boldmath $\nabla$}}

\def\lQ{\Lambda_{\rm QCD}}
\newcommand{\nn}{\nonumber}
\newcommand{\be}{\begin{equation}}
\newcommand{\ee}{\end{equation}}
\newcommand{\bea}{\begin{eqnarray}}
\newcommand{\eea}{\end{eqnarray}}

\def\als{\alpha_{\rm s}}
\def\siml{{\ \lower-1.2pt\vbox{\hbox{\rlap{$<$}\lower6pt\vbox{\hbox{$\sim$}}}}\ }} 
\def\vbfD{{\ \lower-8pt\vbox{\hbox{\rlap{$\!\leftrightarrow$}\lower8pt\vbox{\hbox{$\!\bf D$}}}}\ }} 

\begin{document}
\begin{frontmatter}
\begin{flushright}
\tt{CERN-TH/2003-148 \\ IFUM-759-FT \\ UB-ECM-PF 03/17}
\end{flushright}
\vskip 1truecm
\title{The $\sqrt{m\lQ}$ scale in heavy quarkonium}
\author {Nora Brambilla$^1$, Antonio Pineda$^2$, Joan Soto$^2$ and Antonio Vairo$^3$}
\address{$^1$ INFN and Dipartimento di Fisica dell'Universit\`a di Milano \\
via Celoria 16, 20133 Milan, Italy}
\address{$^2$ Dept. d'Estructura i Constituents de la Mat\`eria and IFAE, 
     U. Barcelona \\ Diagonal 647, E-08028 Barcelona, Catalonia, Spain}
\address{$^3$ Theory Division CERN, 1211 Geneva 23, Switzerland}

\begin{abstract}
We investigate the effects produced by the three-momentum scale
$\sqrt{m\lQ}$ in the strong coupling regime of heavy quarkonium. We
compute the leading non-vanishing contributions due to this scale to
the masses and inclusive decay widths. We find that they may provide
leading corrections to the $S$-wave decay widths but only subleading 
corrections to the masses.
\end{abstract}

\vspace{1cm}
\end{frontmatter}

\newpage

\section{Introduction}

Heavy quarkonium is characterized by the small relative velocity $v$
of the heavy quarks in their centre-of-mass frame. This small
parameter produces a hierarchy of widely separated scales once
multiplied by the mass $m$ of the heavy particle: $m$ (hard), $mv$
(soft), $mv^2$ (ultrasoft), $\dots$. In general, we have $E\sim mv^2
\ll p \sim mv \ll m$, where $E$ is the binding energy and $p$ the
relative three momentum.

It is usually believed that for most of the heavy quarkonium states a
weak coupling analysis is not reliable. However, one can still exploit
the hierarchy of scales in the problem \cite{BBL}. It was argued in
\cite{pnrqcd,m} that in the particular case $\lQ \gg mv^2$, which we
will be concerned with in this letter, it is possible to encode all
the relevant information of QCD in an effective Schr\"odinger-like
description of these systems.  The problem then reduces to calculating
the potentials from QCD.  It has been shown in \cite{m} how to
systematically calculate the potentials within a $1/m$ expansion (see
\cite{potentials} for earlier calculations).

Once the methodology to compute the potentials within a $1/m$
expansion has been developed, the next question appears naturally: at
which extent one can compute the {\it full} potential within a $1/m$
expansion in the case $\lQ \gg mv^2$.\footnote{In fact, there is at
least one example where powers of $\sqrt{m}$ arise upon integrating
out some non-relativistic degrees of freedom \cite{pionium}.}  We
tackle this issue in this paper. We will see that, indeed, new
non-analytical  terms arise due to the three momentum scale
$\sqrt{m\lQ}$. These terms can be incorporated into local potentials
($\delta^3 ({\bf r})$ and derivatives of it) and scale as half-integer
powers of $1/m$. Moreover, we show that it is possible to factorize
these effects in a model independent way and compute them within a
systematic expansion in some small parameters.

As mentioned before, these terms are due to the existence of degrees
of freedom, namely the quark-antiquark pair, with relative
three momentum of order $\sqrt{m\lQ}$.  The on-shell energy of these
degrees of freedom is of $O(\lQ)$, i.e.  the same energy scale that is
integrated out when computing the standard $1/m$ potentials, which
corresponds to integrating out (off-shell) quark-antiquark pairs of
three momentum of order $\lQ$.  Therefore, in principle, both degrees
of freedom should be integrated out at the same time.

In this letter, under the general condition $\lQ \gg mv^2$,  
we will perform the analysis in two possible cases:\\ \\
$1)$ in Sec. \ref{secA} we will consider the particular case $mv \gg \lQ$;  \\ \\ 
$2)$ in Sec. \ref{secB} the general case $\lQ \siml  mv$. \\ \\ 

Note that the scale $\sqrt{m\lQ}$ fulfils $\sqrt{m\lQ}$ $\gg$ $mv$ and $\sqrt{m\lQ}$ $\gg \lQ$. 
From the last inequality it follows that at this scale we always are in the
weak coupling regime.

\section{Case $mv  \gg \lQ$}
\label{secA}
In the case $mv \gg \lQ$, all quarks and gluons with energy 
much larger than $\lQ$ (in particular gluons with energy and momentum 
of order $\sqrt{m\lQ}$ and $mv$) may be integrated out from NRQCD
using weak coupling techniques. 
This leads to the EFT called pNRQCD$^\prime$ in \cite{pw,sw} 
(formerly called pNRQCD in \cite{Mont,pnrqcd}).
This EFT contains, as explicit degrees of freedom, 
gluons with energy and momentum smaller than $m v$ and 
quarks with energy smaller than $m v$ and momentum smaller than $m$.
Quarks may be arranged in quark-antiquark singlet    
${\rm S} = { S 1\!\!{\rm l}_c /\sqrt{N_c}}$ and octet 
${\rm O} = 1/\sqrt{T_F}\, O^a T^a $ fields ($T_F=1/2$).
The Lagrangian of pNRQCD$^\prime$ then reads (${\bf R}$ is the centre-of-mass
coordinate and ${\bf r}$ the relative coordinate) \cite{pnrqcd}:  
\bea
& & L_{\rm pNRQCD^\prime} =\int d^3 R\, \int d^3 r \, \Biggl(
{\rm Tr} \,\Biggl\{ {\rm S}^\dagger \left( i\partial_0 - h_s \right) {\rm S} 
+ {\rm O}^\dagger \left( iD_0 - h_o \right) 
{\rm O} \Biggr\}
\nn\\
& &\qquad + {\rm Tr} \left\{  {\rm O}^\dagger {\bf r} \cdot g{\bf E} \,{\rm S}
+ {\rm S}^\dagger {\bf r} \cdot {\bf E} \,{\rm O} \right\} 
+ {1\over 2} {\rm Tr} \left\{  {\rm O}^\dagger {\bf r} \cdot g{\bf E} \, {\rm O} 
+ {\rm O}^\dagger {\rm O} {\bf r} \cdot g{\bf E}  \right\} \Biggr) +
\int d^3 R\, {\cal L}_g,
\label{pnrqcd0}
\eea
where $h_s=-\bfnabla^2_r/m + V_s$, $h_o=-\bfnabla^2_r/m + V_o$, 
and ${\cal L}_g$ stays for the Lagrangian density of gluons and light quarks.  
The potentials $V=\{V_s, V_o\}$ contain real and imaginary parts. 
The real part, which at leading order is the Coulomb potential $V^{(0)}$, 
has been calculated by different authors over the past years \cite{pertpot}. 
The imaginary part has been calculated in \cite{pw,sw}. 
It consists of local potentials ($\delta^3({\bf r})$ and derivatives of it).
The imaginary coefficients come from the imaginary parts 
of the four-fermion matching coefficients of NRQCD \cite{BBL}.

The next energy scale to be integrated out is $\lQ$. 
This means integrating out all quarks and gluons of energy 
or kinetic energy of order $\lQ$.
The contributions due to (off shell) heavy quarks of energy $\sim \lQ$ 
and three momentum of order $mv$ or smaller
(i.e. of order $\lQ$) are easily singled out by performing an expansion 
of the incoming and outgoing bound-state energies 
$h_s$ and $h_o$ over  $\lQ$ in the matching calculation.
This ensures that the quark kinetic energy is much smaller 
than $\lQ$ and, therefore, that the quark three-momenta are much smaller than 
$\sqrt{m\lQ}$. This expansion only produces terms that are analytical in
$1/m$ \cite{pw,sw}. 

The contributions due to heavy quarks of three momentum of order
$\sqrt{m\lQ}$ may be obtained as follows.
We split the singlet and octet fields of the pNRQCD$^\prime$ Lagrangian into two fields:
\be
S = S_p + S_{sh},  \qquad\qquad O^a = O_p^a + O_{sh}^a,
\ee
where the semi-hard fields $S_{sh}$ and $O_{sh}^a$ are associated to
three-momentum fluctuations of $O\left(\sqrt{m\lQ}\right)$ and the
potential fields $S_p$ and $O_p^a$ to three-momentum fluctuations of $O(m v)$. The potentials are labeled according to the relative momenta 
that they connect: $V = V^{p,p}+V^{p,sh}+V^{sh,p}+V^{sh,sh}$. 
The typical three-momentum transfer in $V^{p,sh}$, $V^{sh,p}$ and
$V^{sh,sh}$ is $\sqrt{m\lQ}$ ($\gg mv$).

The pNRQCD$^\prime$ Lagrangian then reads  
\be
L_{pNRQCD^\prime}=L_g + L^{sh}_{pNRQCD^\prime}+L^p_{pNRQCD^\prime}+L_{mixing}.
\ee
The expressions for $L^{sh}_{pNRQCD^\prime}$ and $L^p_{pNRQCD^\prime}$ are identical 
to the pNRQCD$^\prime$ Lagrangian except for the changes 
$S$, $O^a$, $V_s$, $V_o$  $\rightarrow$ $S_{sh}$, $O_{sh}^a$, $V_s^{sh,sh}$, $V_o^{sh,sh}$ 
and $S$, $O^a$,
$V_s$, $V_o$  $\rightarrow$ $S_{p}$, $O_{p}^a$, $V_s^{p,p}$, $V_o^{p,p}$   respectively.
Recall that the gluons left dynamical are of $O(\lQ)$ and that 
analytical  terms in ${\bf r}$ do not mix semi-hard and potential fields. 
Therefore, the multipole expansion in (\ref{pnrqcd0}) 
is an expansion with respect to either the scale ${\bf r}\sim
1/\sqrt{m\lQ}$ in $L^{sh}_{pNRQCD^\prime}$ or the scale ${\bf r}\sim 1/mv$ 
in $L^p_{pNRQCD^\prime}$. 

Throughout the paper we will also assume that 
\be
\sqrt{m\lQ} \gg m\,\als(\sqrt{m\lQ}) \, , 
\label{count}
\ee
which implies that the
Coulomb potentials in $V^{p,sh}$, $V^{sh,p}$ and $V^{sh,sh}$ can be
expanded about the kinetic energy and no Coulomb resummation is needed. This is not so for $ V^{p,p}$.  

The leading contribution to the real part of $L_{mixing}$ comes from the mixing of 
$S_{sh}$ with $S_p$ and $O_{sh}^a$ with $O_p^a$ due to the Coulomb potential. 
As an example, consider the real part of the singlet-mixing term
due to the static Coulomb potential. It is given by 
\bea
\label{exp}
&&{\rm Re} \, L_{mixing}\Bigg|_{\rm Singlet} = 
-\int d^3R\,\int d^3r \, 
S^\dagger_p({\bf R},{\bf r})\, V^{(0)}_s({\bf r}) \, S_{sh}({\bf R},{\bf r}) 
+ {\rm H.c.} 
\\
&&
\qquad
= 
-\int d^3R\,\int d^3p \, \int d^3p'
\tilde{S}^\dagger_{p}({\bf R},{\bf p}) \, \tilde{V}^{(0)}_s({\bf p}-{\bf p}') \, 
\tilde{S}_{sh}({\bf R},{\bf p}')
+ {\rm H.c.} 
\nn \\
\nn
&&
\qquad
= 
-\int d^3R\,\int d^3r\,
\left(
S^\dagger_{p}({\bf R},{\bf 0})+{\bf r}\cdot \bfnabla_{\bf
r}S^\dagger_{p}({\bf R},{\bf r})\Bigg|_{{\bf r}={\bf 0}}+\cdots 
\right)
\, V^{(0)}_s({\bf r}) \, 
S_{sh}({\bf R},{\bf r})
+ {\rm H.c.}\;.
\eea
In order to avoid a cumbersome notation we have dropped the upper-index
${p,sh}$ from $V^{(0)}_s({\bf r})$. In fact, any potential between
fields labeled by $a,b=p,sh$ always has upper-indices ${a,b}$. Hence,
dropping the upper-indices shall not lead to ambiguities.
In the second line of Eq. (\ref{exp}), a
Fourier transform of all the fields has been performed, and in the
third one, we have expanded around ${\bf p} \sim 0$ in the potential,
since, by definition, ${\bf p} \sim m v \ll {\bf p}' \sim
\sqrt{m\lQ}$.  Doing so in the loops that will appear in the matching
computation guarantees that only the scale $\sqrt{m\lQ}$ is integrated
out. Alternatively, one may consider $S^\dagger_p({\bf R},{\bf r})$ 
slowly varying in ${\bf r}$ and multipole expand it about ${\bf r} = {\bf 0}$, 
which brings us directly from the first to the last line of Eq. (\ref{exp}).
At the order of interest we have $V^{(0)}_s = -C_f\, \als/r$ and
$\als=\als\left(\sqrt{m\lQ}\right)$.  Analogous results hold for the
real part of the octet-mixing term due to the static Coulomb
potential:
\bea
&&
{\rm Re} \, L_{mixing}\Bigg|_{\rm Octet} 
\\
\nn
&&
= 
-\int d^3R\,\int d^3r\, {\rm Tr} \left\{ 
\left(
{\rm O}^\dagger_{p}({\bf R},{\bf 0}) 
+{\bf r}\cdot \bfnabla_{\bf
r}O^\dagger_{p}({\bf R},{\bf r})\Bigg|_{{\bf r}={\bf 0}}+\cdots
\right)
\, V^{(0)}_o({\bf r}) \, 
{\rm O}_{sh}({\bf R},{\bf r}) 
+ {\rm H.c.} \right\},
\eea
where the trace is over the colour indices, the mixing potential is $V^{(0)}_o = 1/(2N_c)\,\als/r$ and
$\als=\als\left(\sqrt{m\lQ}\right)$.

The leading contribution to the imaginary part of $L_{mixing}$ can be immediately 
read off from the imaginary delta-type potentials calculated in \cite{sw}:
\bea
{\rm Im} \, L_{mixing} &=& - \int d^3R\,\int d^3r\,  {\rm Tr} \, \left\{ 
{\rm S}^\dagger_{sh}({\bf R},{\bf 0}) \, 
{K_s\over m^2} \delta^3({\bf r})  \, {\rm S}_p({\bf R},{\bf 0})  + {\rm H.c.} \right\}  
\nn\\
&& - \int d^3R\,\int d^3r\,  {\rm Tr} \, \left\{ 
{\rm O}^\dagger_{sh}({\bf R},{\bf 0}) \, 
{K_o\over m^2} \delta^3({\bf r}) \, {\rm O}_p({\bf R},{\bf 0})  + {\rm H.c.}  \right\} , 
\label{impot}
\eea
where 
\bea
K_s&=&-{C_A \over 2}
\Bigg(
4\, {\rm Im} \, f_1(^1 S_0)
-2\,{\bf S}^2\left({\rm Im}\, f_1(^1 S_0)-{\rm Im}\, f_1(^3 S_1)\right)
\nn \\
&&
\qquad\qquad
+ 4\,{\rm Im}\, f_{\rm EM}(^1 S_0)
-2\,{\bf S}^2\left({\rm Im} \, f_{\rm EM}(^1 S_0)-{\rm Im}\, f_{\rm EM}(^3 S_1)\right)
\Bigg), 
\label{K}
\\
K_o&=&-{T_F\over 2}
\Bigg(
4\, {\rm Im} \, f_8(^1 S_0)
-2\,{\bf S}^2\left({\rm Im}\, f_8(^1 S_0)-{\rm Im}\, f_8(^3 S_1)\right) \Bigg).
\eea
The matching coefficients $f$ are the matching coefficients of the
four-fermion operators in NRQCD and may be read off from Ref. \cite{BBL}.

\subsection{Matching}
\label{matching1}
The next step is to integrate out from pNRQCD$^\prime$ all
fluctuations that appear at the energy scale $\lQ$. 
These are light quarks and gluons of energy or three momentum  of
order $\lQ$, and singlet and octet fields of energy of order $\lQ$ or
three momentum of order $\sqrt{m\lQ}$.
We will be left with pNRQCD, where only a singlet field
describing a quark-antiquark pair of energy $mv^2$ and relative
three-momentum $mv$ is dynamical\footnote{We ignore pseudo-Goldstone bosons (pions), which, in principle, should also be included.}:
\bea
L_{pNRQCD^\prime} \rightarrow L_{pNRQCD} & =& L_{pNRQCD}^{1/m} + L_{pNRQCD}^{1/\sqrt{m}}\,,\\
 L_{pNRQCD}^{1/m}&=& \int d^3R\int d^3r \, S^\dagger \, \left( i\partial_0-{{\bf p}^2\over m}-V_s^{p,p}-\delta V^{1/m}\right) \, S\,,  \\
L_{pNRQCD}^{1/\sqrt{m}} &=&-\int d^3R\int d^3r \, S^\dagger \, \delta V^{1/\sqrt{m}} \, S
\,.
\eea
$L_{pNRQCD}^{1/m}$ is defined as the part of the pNRQCD Lagrangian obtained by
integrating out quarks and gluons of energy and three-momentum of
order $\lQ$ in $L_{pNRQCD^\prime}^p$ only.  It is analytical  in $1/m$ and has
been considered before in \cite{pnrqcd,sw}.  Here we will calculate
the leading part of $L_{pNRQCD}^{1/\sqrt{m}}$, which is defined as the part of the pNRQCD Lagrangian
obtained by integrating out quark-antiquark pairs of three-momentum
$\sqrt{m\lQ}$ in $L_{pNRQCD^\prime}$ in addition to the above degrees of
freedom. In general, it is non analytical  in $1/m$, and, at leading order,
it consists of a new local (delta-type) potential.

\begin{figure}[htb]
\vskip 0.8truecm
\makebox[0truecm]{\phantom b}
\put(10,0){\epsfxsize=6.5truecm\epsffile{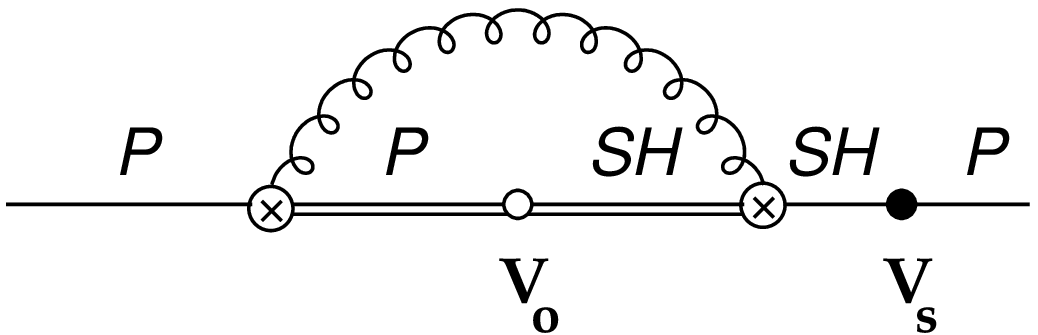}}
\put(5,60){\it a)}
\put(250,0){\epsfxsize=6.5truecm\epsffile{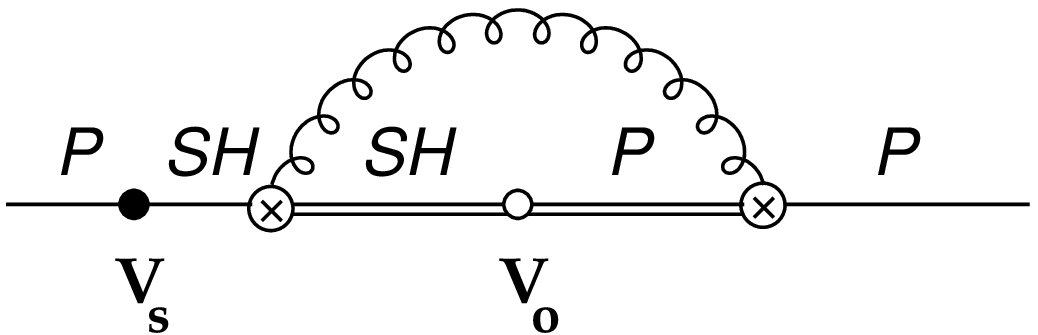}}
\put(245,60){\it b)}
\put(10,-100){\epsfxsize=6.5truecm\epsffile{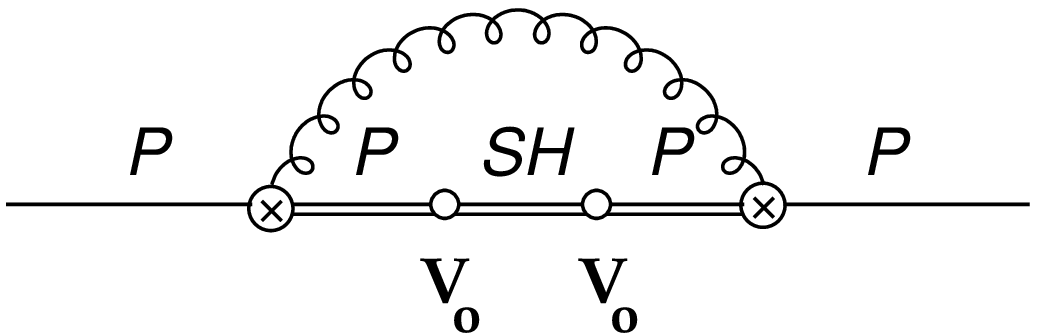}}
\put(5,-40){\it c)}
\put(250,-100){\epsfxsize=6.5truecm\epsffile{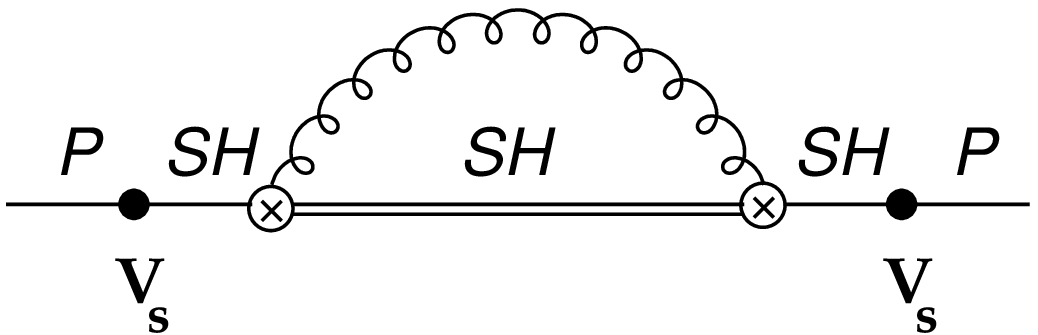}}
\put(245,-40){\it d)}
\caption{\it The four diagrams of pNRQCD$^\prime$ contributing to 
$\delta V^{1/\sqrt{m}}$ at leading (non-vanishing) order in the multipole expansion. 
Open and full circles indicate 
octet and singlet potential insertions coming from the mixing terms 
respectively. These are treated according to Eq. (\ref{exp}).
The upper-scripts $P$ and $SH$ on a propagator indicate that the propagating
fields are of the potential and semi-hard type respectively. The circle with a
cross indicates the vertex ${\rm S}^\dagger \, {\bf r}\cdot {\bf E} \, {\rm
  O}$ (or Hermitian conjugate), where the 
quark fields are both either potential or semi-hard. The gluon line stands for 
non-perturbative multi-gluon exchanges.}
\label{figen}
\end{figure}
The matching condition for the full $\delta V=\delta V^{1/m}+\delta V^{1/\sqrt{m}}$ 
at leading (non-vanishing) order in the multipole expansion is 
\bea
&& {1\over E-{{\bf p}^2\over m}-V_s^{p,p}}\delta V{1\over E-{{\bf p}^2\over
    m}-V_s^{p,p}}=
\nn\\
&&\qquad\qquad\qquad\qquad
{1 \over E-h_s} \, {i \over N_c}  \int_0^{\infty} dt \, \langle {\rm vac} | i {\bf r}
\cdot g{\bf E} (t) \, e^{-i(h_o-E)t} \, i {\bf r} \cdot g{\bf E} (0) 
|{\rm vac}\rangle \, {1\over
  E-h_s} \, .
\label{matching}
\eea
The above matching equation should be understood (even if written at
the operator level) with incoming (outcoming) momenta and energy $E$ of
$O(mv)$ and $O(mv^2)$ respectively. 
The typical size of the time variable in the integral is given by the vacuum expectation value 
of the chromoelectric correlator and hence $t \sim 1/\lQ$. 
The separation between potential and semi-hard relative three-momenta discussed
above can be easily implemented in the rhs of Eq. (\ref{matching}) by expanding 
the Hamiltonians $h_{s,o}$ in $V^{sh,p}_{s,o}$ and $V^{p,sh}_{s,o}$. 
The zeroth order term in this expansion gives $\delta V^{1/m}$ and has been calculated 
in \cite{pnrqcd,sw}. The $V^{p,p}$ potential cannot be expanded in the potential region.
The size of the three-momenta in the semi-hard regions is of $O(\sqrt{m\lQ})$. 
Several approximations apply: 
(i) $E-h_{o,s}\sim -{\bf p}^2/ m \sim \lQ$ in the semi-hard regions,
(ii) $(h_o-E)t \sim ({{\bf p}^2\over m}+V_o^{p,p}-E)t \sim mv^2/\lQ
\ll 1 $ in the potential regions
and (iii) we can expand the potential three-momenta with respect to the semi-hard 
ones in $ V_{s,o}^{sh,p}$ and $ V_{s,o}^{p,sh}$. 

The leading contributions to  $\delta V^{1/\sqrt{m}}$ have been 
depicted in Fig. \ref{figen}. Fig. \ref{figen}a corresponds to
\be
\delta V^{1/\sqrt{m}}= {i \over N_c}  \int_0^{\infty} dt \,  \int_0^{t} dt^{\prime}
\, \langle {\rm vac}| i {\bf r}
\cdot g{\bf E} (t) \, 
\left( -iV_o^{p,sh}\right) e^{-i{{\bf p}^2\over m}t^{\prime}} 
i {\bf r} \cdot g{\bf E} (0) 
|{\rm vac}\rangle {1\over {-{\bf p}^2\over m}} V_s^{sh,p},
\label{figa}
\ee
Fig. \ref{figen}b corresponds to
\be
\hspace{-5mm}
\delta V^{1/\sqrt{m}}= V_s^{p,sh}{1\over {-{\bf p}^2\over m}}{i \over N_c}  
\int_0^{\infty} dt \,  \int_0^{t} dt^{\prime} \, 
\langle {\rm vac}| i {\bf r}
\cdot g{\bf E} (t) \, e^{-i{{\bf p}^2\over m}(t-t^{\prime})} \, 
\left( -iV_o^{sh,p}\right) 
\,
i {\bf r} \cdot g{\bf E} (0) 
|{\rm vac}\rangle,
\label{figb}
\ee
Fig. \ref{figen}c corresponds to
\bea
\delta V^{1/\sqrt{m}}&=& {i \over N_c}  \int_0^{\infty} dt \,  
\int_0^{t} dt^{\prime} \,\int_0^{t^\prime} dt^{\prime\prime} \, 
\langle {\rm vac}| i {\bf r}
\cdot g{\bf E} (t) \, 
\nn \\ & & \qquad\qquad
\times \left( -iV_o^{p,sh}\right)e^{-i{{\bf p}^2\over
    m}(t^{\prime}-t^{\prime\prime})} 
\left( -iV_o^{sh,p}\right)
i {\bf r} \cdot g{\bf E} (0) 
|{\rm vac}\rangle,
\label{figc}
\eea
and, finally, Fig. \ref{figen}d corresponds to
\bea
\delta V^{1/\sqrt{m}}&=& V_s^{p,sh}{1\over {-{\bf p}^2\over m}}{i \over N_c}  
\int_0^{\infty} dt  \, 
\langle {\rm vac}| i {\bf r}
\cdot g{\bf E} (t) \, e^{-i{{\bf p}^2\over m}t} \, 
 i {\bf r} \cdot g{\bf E} (0) 
|{\rm vac}\rangle {1\over {-{\bf p}^2\over m}} V_s^{sh,p}.
\eea
  
The potential $\delta V$ contains a real and an imaginary part.  The
real part contributes to the heavy quarkonium spectrum, the imaginary
one to the inclusive decay width.

\subsection{Corrections to the spectrum}
The four diagrams that give the leading contribution to ${\rm Re}\, \delta V^{1/\sqrt{m}}$
are obtained from those of Fig. \ref{figen} by substituting
$V_{s,o}\rightarrow V_{s,o}^{(0)}$,  
where $V_{s,o}^{(0)}$ are the Coulomb singlet and octet potentials. They give:
\bea
{\rm Re}\, \delta V^{1/\sqrt{m}}
&=&
-i^{9/2} (2C_f+C_A)^2 {64 \over 315} 
\sqrt{\pi}\, \als^2 \, {\cal E}_{7/2} \; {\delta^3({\bf r}) \over m^{3/2}}
\nn\\
&=&
(2C_f+C_A)^2{4 \over 3\Gamma(9/2)}
\pi \, \als^2 \, {\cal E}_{7/2}^E \; {\delta^3({\bf r}) \over m^{3/2}}\,,
\label{ReVpert}
\eea
where in the first equality we have used the definition of ${\cal E}_{n}$ that one may find in 
Ref. \cite{sw} and in the last equality we have written the chromoelectric correlator in 
Euclidean space (traces as well as 
suitable Schwinger lines connecting the gluon fields are understood):
\be
{\cal E}_n^E
=
{1 \over N_c}\int_0^\infty d\tau\, \tau^{n} \langle {\rm vac}|
g{\bf E}(\tau)\cdot g{\bf E}(0) |{\rm vac}\rangle_E
\,.
\ee 

Eq. (\ref{ReVpert}) gives a contribution to the energy of $\displaystyle O\left(mv^3\als \,
{m\als \over \sqrt{m\lQ}}\right)$. 

\subsection{Corrections to the decay width}
\label{secdec}
\begin{figure}[htb]
\vskip 0.8truecm
\makebox[0truecm]{\phantom b}
\put(10,0){\epsfxsize=6.5truecm\epsffile{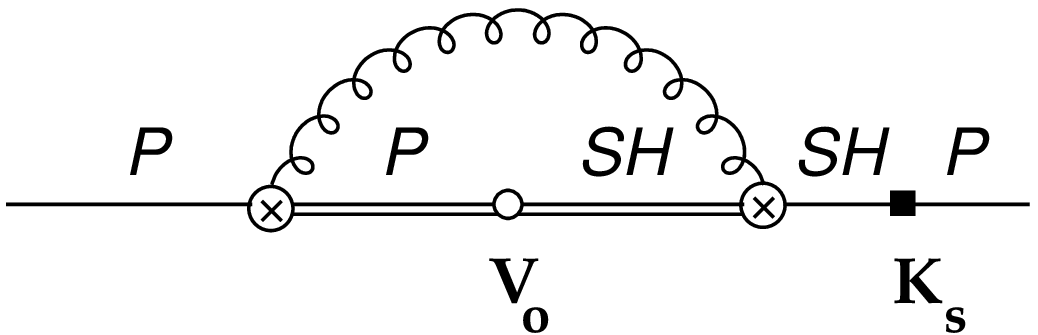}}
\put(5,60){\it a)}
\put(250,0){\epsfxsize=6.5truecm\epsffile{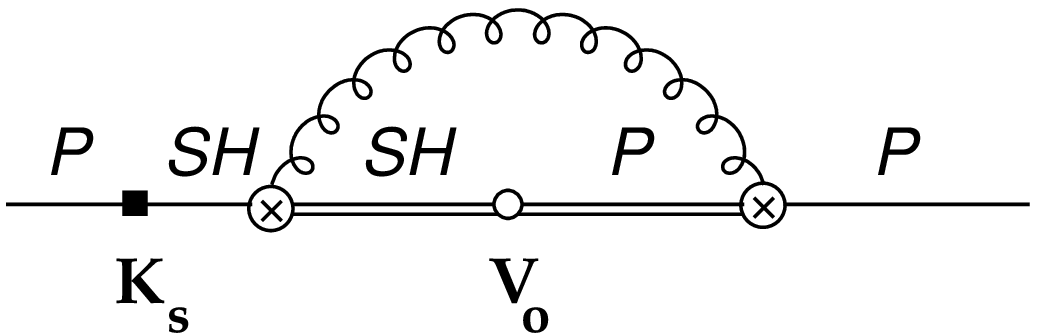}}
\put(245,60){\it b)}
\put(10,-100){\epsfxsize=6.5truecm\epsffile{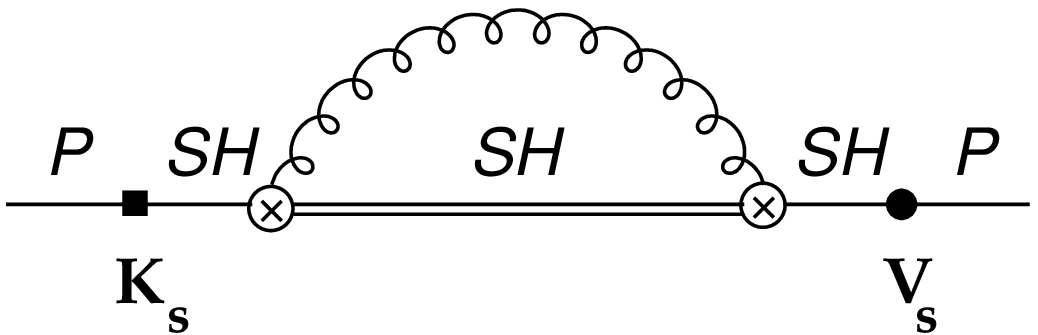}}
\put(5,-40){\it c)}
\put(250,-100){\epsfxsize=6.5truecm\epsffile{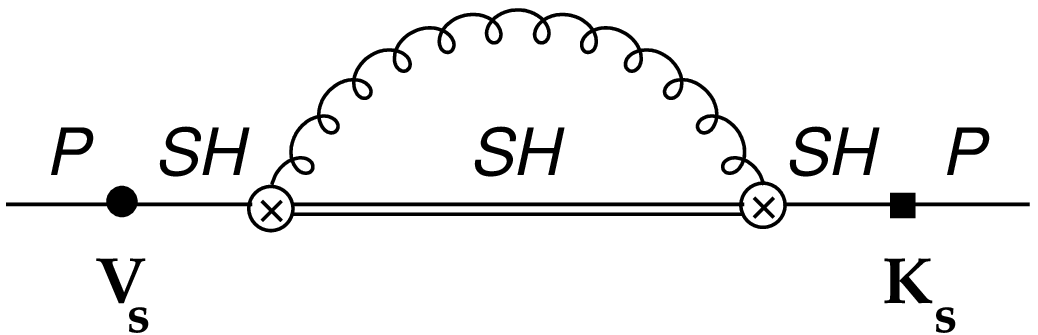}}
\put(245,-40){\it d)}
\caption{\it The four diagrams of pNRQCD$^\prime$ contributing to 
${\rm Im}\,\delta V^{1/\sqrt{m}}$ at the leading (non-vanishing) order in the multipole expansion. 
The full box indicates the insertion of a delta-type potential proportional to $K_s$. All other 
symbols are as in Fig. \ref{figen} with $V_{s,o}\rightarrow V_{s,o}^{(0)}$.}
\label{figde}
\end{figure}

The four diagrams that give the leading contribution to ${\rm Im}\, \delta V^{1/\sqrt{m}}$
are shown in Fig. \ref{figde}. These can be derived  from the diagrams of 
Fig. \ref{figen} by replacing one of the potentials by a Coulomb potential and the second potential
 with the imaginary delta potential of Eq. (\ref{impot}). The graph with 
two potentials inside the gluonic loop as well as graphs 
involving the octet delta potential ($\sim K_o \, \delta^3({\bf r})/m^2$) 
do not contribute to ${\rm Im}\, \delta V$ as a delta potential (although they
do as derivatives of a delta potential, which are subleading). 
We obtain
\bea
{\rm Im} \, \delta V^{sh}
&=&
-i^{7/2} {32 \over 45} \, (2C_f+C_A)
{1\over \sqrt{\pi}} \, K_s\, \als\, {\cal E}_{5/2} \; {\delta^3({\bf r})\over m^{5/2}}
\nn\\
&=&
(2C_f+C_A) {4 \over 3\Gamma(7/2)} \, K_s\, \als\,
 {\cal E}^E_{5/2}\; {\delta^3({\bf r})\over m^{5/2}} \,,
\eea
where in the last equality we have written the chromoelectric correlator in 
Euclidean space.

The above correction gives a contribution to the decay width suppressed by a
factor of \\ $\displaystyle O\left({\lQ \over m} \, 
{m\als \over \sqrt{m\lQ}}\right)$ with respect to the leading contribution.

A similar analysis can be done for the $P$-wave decays. 
The leading effect would be in that case at least $O(m\als/\sqrt{m\lQ})$ 
suppressed with respect to the leading contribution computed in \cite{pw}.

\section{Case $\lQ \siml mv$}
\label{secB}
Here we will follow the same procedure as in the previous section.
In this case, however, the starting point is the NRQCD Lagrangian. 
We split the quark (antiquark) field into two: a semi-hard field for the
(three-momentum) fluctuations of $O(\sqrt{m\lQ})$, $\psi_{sh}$
($\chi_{sh}$), and a potential field for the (three-momentum)
fluctuations of $O(mv)$, $\psi_p$ ($\chi_{p}$):
\be 
\psi = \psi_p + \psi_{sh}, \qquad \qquad \chi = \chi_p + \chi_{sh}.
\ee
The NRQCD Lagrangian then reads
\be
L_{NRQCD}=L^{sh}_{NRQCD}+L^p_{NRQCD}+L_{mixing}+L_g.
\ee

The Lagrangians $L^{sh}_{NRQCD}$ and $L^p_{NRQCD}$ are identical to the NRQCD 
Lagrangian expressed in terms of semi-hard and potential fields respectively.
The quantity $L_g$ is the QCD Lagrangian for gluons and light quarks. 
For $L^{sh}_{NRQCD}$ we can use weak coupling techniques.  Therefore, we can construct a
pNRQCD$^\prime$ Lagrangian for it, once gluons and quarks of energy or
three momentum of $O(\sqrt{m\lQ})$ have been integrated out and
transformed into potentials:
\be 
L^{sh}_{NRQCD} \to L^{sh}_{pNRQCD'}.
\ee
If we further project to the quark-antiquark sector, the Lagrangian
$L^{sh}_{pNRQCD'}$ will formally read equal to Eq. (\ref{pnrqcd0}). The multipole expanded gluons in
$L^{sh}_{pNRQCD'}$ have (four) momentum much smaller than
$\sqrt{m\lQ}$. We note that we cannot do the same for $L^p_{NRQCD}$ since at scales of
$O(\lQ)$ we can neither use weak coupling techniques nor the multipole
expansion. 

\begin{figure}[htb]
\vskip 0.8truecm
\makebox[0truecm]{\phantom b}
\put(150,0){\epsfxsize=6truecm\epsffile{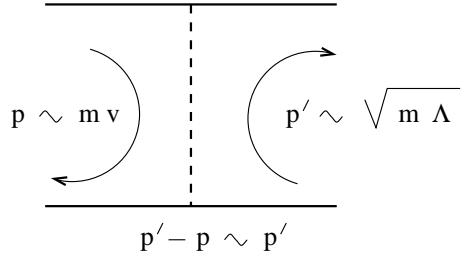}}
\caption{\it The Coulomb-exchange graph contributing to the leading mixing 
interaction between semi-hard and potential fields.}
\label{figmix}
\end{figure}
We consider now $L_{mixing}$. We will assume, as in Sec. \ref{secA}, 
that the condition (\ref{count}) holds. This will allow us to 
treat the Coulomb potential as a perturbation at the semi-hard scale.
The leading order contribution to the real part of $L_{mixing}$ 
comes from the one-Coulomb exchange graph (see Fig. \ref{figmix}):
\bea 
{\rm Re} \, L_{mixing}^{(0)} &=& - \int d^3R \, \int d^3r \, 
{\rm Tr}\left\{ 
J^\dagger({\bf R})
\, V_s^{(0)}({\bf r}) \, {\rm S}_{sh}({\bf R},{\bf r}) \right\} + {\rm H.c.}
\nn\\
&& 
- \int d^3R \, \int d^3r \,
{\rm Tr}\left\{ 
J^\dagger({\bf R})
\, V_o^{(0)}({\bf r}) \, {\rm O}_{sh}({\bf R},{\bf r}) \right\} + {\rm H.c.}\;,
\label{mix0}
\\
J^\dagger({\bf R}) &\equiv& \chi_p({\bf R}) \psi^\dagger_p({\bf R}).
\eea
The potentials $V_s^{(0)}$ and $V_o^{(0)}$ are perturbative:
$V^{(0)}_s = -C_f\, \als/r$ and  $V^{(0)}_o = 1/(2N_c)\,\als/r$.
The coupling constant is calculated at the semi-hard scale $\sqrt{m\lQ}$.
Besides the above term we need to consider also the next-to-leading 
term in the $mv/\sqrt{m\lQ}$ expansion. It is given by
\bea 
{\rm Re} \, L_{mixing}^{(1)} &=&  
-
\int d^3R \, \int d^3r \,
{\rm Tr}\left\{ 
{\bf J}^\dagger({\bf R})\cdot {\bf r}
\, V_s^{(0)}({\bf r}) \, {\rm S}_{sh}({\bf R},{\bf r}) \right\} + {\rm H.c.}
\nn\\
&& 
- \int d^3R \, \int d^3r \,
{\rm Tr}\left\{ 
{\bf J}^\dagger({\bf R})\cdot {\bf r}
\, V_o^{(0)}({\bf r}) \, {\rm O}_{sh}({\bf R},{\bf r}) \right\} + {\rm H.c.}
\label{mix1}
\;,
\\
{\bf J}^\dagger({\bf R}) &\equiv& \chi_p({\bf R}) {\vbfD \over 2} \psi^\dagger_p({\bf R}).
\eea
A practical way to obtain ${\rm Re} \, L_{mixing}^{(1)}$ is by expanding the
Coulomb potential in Fig. \ref{figmix} at higher order in ${\bf p}/{\bf p}'$ and promoting the 
conventional derivatives acting on the potential fields to covariant 
ones. A proper tree-level matching in coordinate space can be done using the field 
redefinitions of Ref. \cite{Mont} for the semi-hard fields projected 
to the two particle sector and multipole expanding the potential fields.
The leading contribution to the imaginary part of $L_{mixing}$ is analogous to
the one given by Eq. (\ref{impot}):
\bea
{\rm Im} \, L_{mixing}^{(0)} &=& 
- \int d^3R\,\int d^3r\,  
{\rm Tr}\left\{ {\rm S}^\dagger_{sh}({\bf R},{\bf 0}) \, 
{K_s\over m^2} \delta^3({\bf r})  \, 
  J({\bf R}) \right\} + {\rm H.c.}  
\nn\\
&& - \int d^3R\,\int d^3r\, {\rm Tr}\left\{ {\rm O}^\dagger_{sh}({\bf R},{\bf 0}) \, 
{K_o\over m^2} \delta^3({\bf r}) \, 
 J({\bf R})\right\}  + {\rm H.c.}  
\label{impotnp}
\;.
\eea  
Note that the potential fields always appear as local currents in
$L_{mixing}$. Finally, the effective field theory that we obtain, at the order of interest, is given by
\be
L_{NRQCD'}=L^{sh}_{pNRQCD'}+L^p_{NRQCD}+
{\rm Re} \, L_{mixing}^{(0)} +  {\rm Re} \, L_{mixing}^{(1)} + 
{\rm Im} \, L_{mixing}^{(0)} + L_g\,,
\ee
where $L_g$ contains now gluons and light quarks of energy and momentum 
much smaller than $\sqrt{m\lQ}$.

\subsection{Matching}
As in section (\ref{matching1}), we now want to integrate out degrees 
of freedom of $O(\lQ)$. We will be left with an EFT, pNRQCD, where only 
a singlet field describing a quark-antiquark pair of energy 
$mv^2$ and relative three momentum $mv$ is dynamical: 
\be
L_{NRQCD^\prime} \rightarrow L_{pNRQCD} = L_{pNRQCD}^{1/m} + L_{pNRQCD}^{1/\sqrt{m}}.
\ee

The quantity $L_{pNRQCD}^{1/m}$ is obtained by integrating out quarks and gluons 
of energy and three momentum of order $\lQ$ in $L_{NRQCD}^{p}$.  It is
analytical 
in $1/m$ and has been considered before in \cite{m,pw,sw}.  
Here we will calculate the leading part of $L_{pNRQCD}^{1/\sqrt{m}}$, which, in
general, is non analytical  in $1/m$. It involves the integration from
NRQCD of quark-antiquark pairs of three momentum
$\sqrt{m\lQ}$. 
The Lagrangian $L_{pNRQCD}^{1/\sqrt{m}}$ will consist, at leading order, of a new local
(delta-type) potential that we name $\delta V^{1/\sqrt{m}}$:
\be
L_{pNRQCD}^{1/\sqrt{m}} = -\int d^3R\int d^3r \, S^\dagger \, \delta V^{1/\sqrt{m}} \, S \,.
\ee 

\begin{figure}[htb]
\vskip 0.8truecm
\makebox[0truecm]{\phantom b}
\put(10,0){\epsfxsize=6.5truecm\epsffile{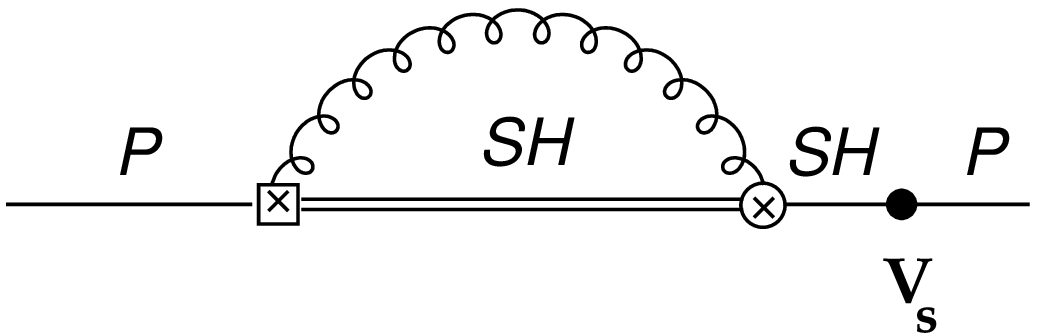}}
\put(5,60){\it a)}
\put(250,0){\epsfxsize=6.5truecm\epsffile{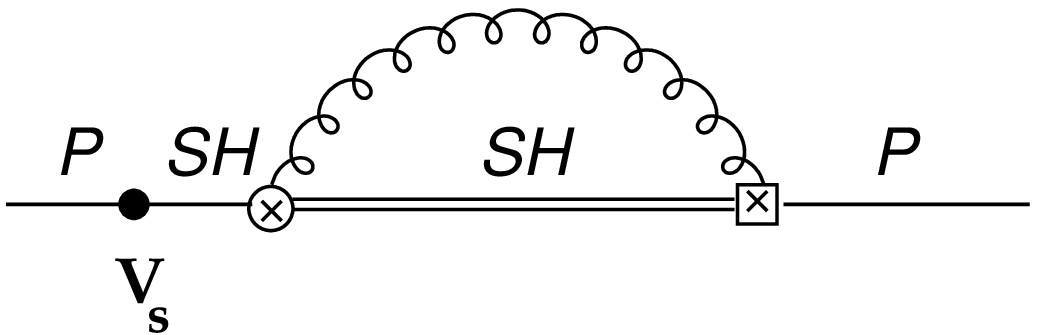}}
\put(245,60){\it b)}
\put(10,-82){\epsfxsize=6.5truecm\epsffile{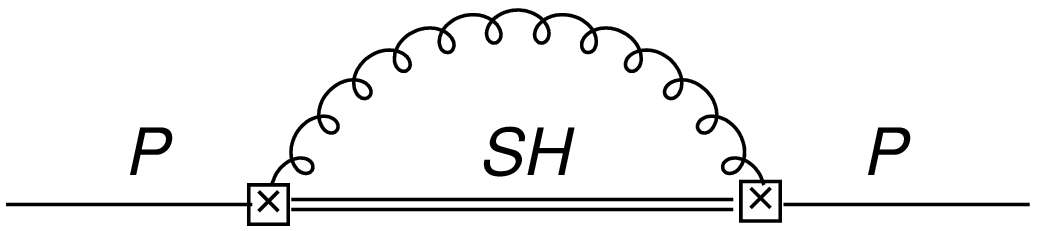}}
\put(5,-40){\it c)}
\put(250,-100){\epsfxsize=6.5truecm\epsffile{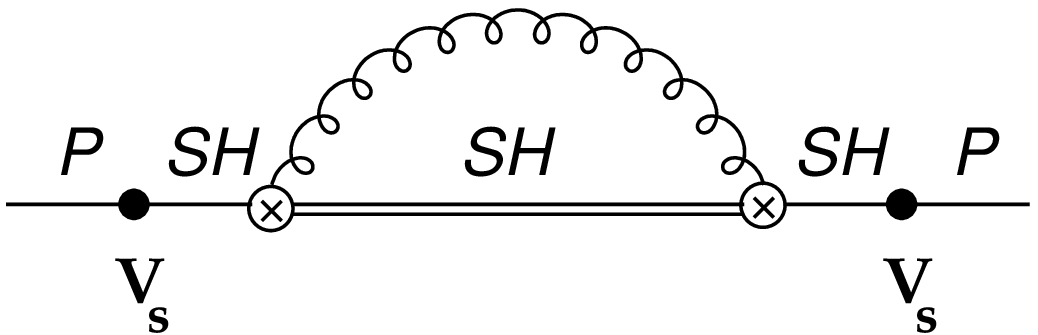}}
\put(245,-40){\it d)}
\caption{\it The four diagrams of NRQCD$^\prime$ contributing to 
$\delta V^{1/\sqrt{m}}$ at leading order. Full circles indicate 
singlet potential insertions coming from ${\rm Re} \, L_{mixing}^{(0)}$.
The upper-scripts $P$ and $SH$ on a propagator indicate that the propagating
fields are of the potential and semi-hard type respectively. The circle with a
cross indicates the vertex ${\rm S}^\dagger_{sh} \, {\bf r}\cdot {\bf
E} \, {\rm O}_{sh}$ (or Hermitian conjugate).  
The box with a cross indicates the vertex
${\rm Re} \, L_{mixing}^{(1)}$. The gluon line symbolizes 
multigluon non-perturbative exchanges.}
\label{figen2}
\end{figure}

The matching calculation for $\delta V^{1/\sqrt{m}}$ is analogous to the
computation of the previous section supplemented with the
technology developed in Refs.  \cite{m,pw,sw}. The leading contribution 
is given by the four diagrams shown in Fig. \ref{figen2}. 
Fig. \ref{figen2}a corresponds to (according to the notation of Ref. \cite{m}) 
\bea 
&& \delta V^{1/\sqrt{m}} ({\bf x}_1-{\bf x}_2) \delta^3({\bf x}_1-{\bf x}_1^\prime)
\delta^3({\bf x}_2-{\bf x}_2^\prime) =
\int d^3R\;\; 
{}^{(0)}\langle \underline{0}; {\bf x}_1, {\bf x}_2| 
\, {\bf J}^i({\bf R}) \, 
\nn\\
&& \times 
\left[
\langle {\bf p}={\bf 0}| \, 
{\bf r}^i \, V_o^{(0)}(r) \, 
{1 \over -{ {\bf p}^2 \over m} -H}{\bf r}^j 
{1 \over -{ {\bf p}^2 \over m}} \, 
V_s^{(0)}(r) \, 
|{\bf p}={\bf 0} \rangle
\right] \, 
g {\bf E}^j({\bf R}) \, J({\bf R}) \, 
|\underline{0}; {\bf x}_1^\prime, {\bf x}_2^\prime \rangle^{(0)}
\,, 
\eea 
Fig. \ref{figen2}b corresponds to 
\bea 
&& \delta V^{1/\sqrt{m}} ({\bf x}_1-{\bf x}_2) \delta^3({\bf x}_1-{\bf x}_1^\prime)
\delta^3({\bf x}_2-{\bf x}_2^\prime) =
\int d^3R\;\;
{}^{(0)}\langle \underline{0}; {\bf x}_1, {\bf x}_2| \, 
g {\bf E}^i({\bf R}) \, J({\bf R}) \, 
\nn\\ 
&& \times 
\left[
\langle {\bf p}={\bf 0}|\, 
V_s^{(0)}(r) \, 
{1 \over -{ {\bf p}^2 \over m}} \,
{\bf r}^i \,
{1 \over -{ {\bf p}^2 \over m} -H} \, 
{\bf r}^j \, V_o^{(0)}(r) \, 
|{\bf p}={\bf 0} \rangle
\right] \, 
{\bf J}^j({\bf R})\, 
|\underline{0}; {\bf x}_1^\prime, {\bf x}_2^\prime \rangle^{(0)}
\,, 
\eea
Fig. \ref{figen2}c corresponds to 
\bea 
&& \delta V^{1/\sqrt{m}} ({\bf x}_1-{\bf x}_2) \delta^3({\bf x}_1-{\bf x}_1^\prime)
\delta^3({\bf x}_2-{\bf x}_2^\prime) =
\int d^3R\;\; 
{}^{(0)}\langle \underline{0}; {\bf x}_1, {\bf x}_2| \, 
{\bf J}^i({\bf R}) \, 
\nn\\
&& \times 
\left[
\langle {\bf p}={\bf 0}| \,
{\bf r}^i \, V_o^{(0)}(r)  \, 
{{1 \over -{ {\bf p}^2 \over m} -H}} \, 
{\bf r}^j \, V_o^{(0)}(r) \,  
|{\bf p}={\bf 0} \rangle
\right] \, 
{\bf J}^j({\bf R}) \, 
|\underline{0}; {\bf x}_1^\prime, {\bf x}_2^\prime \rangle^{(0)}
\,.
\eea 
Finally, Fig. \ref{figen2}d gives 
\bea 
&& \delta V^{1/\sqrt{m}} ({\bf x}_1-{\bf x}_2) \delta^3({\bf x}_1-{\bf x}_1^\prime)
\delta^3({\bf x}_2-{\bf x}_2^\prime) =
\int d^3R\;\; 
{}^{(0)}\langle \underline{0}; {\bf x}_1, {\bf x}_2| \, 
g {\bf E}^i({\bf R}) \, J({\bf R}) \, 
\nn\\
&& \times
\left[
\langle {\bf p}={\bf 0}|\, 
V_s^{(0)}(r) \, 
{1 \over -{ {\bf p}^2 \over m}} \,
{\bf r}^i \,
{1 \over -{ {\bf p}^2 \over m} -H} \, 
{\bf r}^j \, 
{1 \over -{ {\bf p}^2 \over m}} \,
V_s^{(0)}(r) \, 
|{\bf p}={\bf 0} \rangle
\right] \, 
\nn\\
&& \times \, 
g {\bf E}^j({\bf R}) \, J({\bf R}) \, 
|\underline{0}; {\bf x}_1^\prime, {\bf x}_2^\prime \rangle^{(0)}
\,.
\eea 
In the above equations $H$ stands for the NRQCD Hamiltonian in the
static limit and $| 0; {\bf r} \rangle^{(0)}$ is the gluonic piece of the
ground state of NRQCD in the static limit. We refer to \cite{m,sw} for
further details. As in Sec. \ref{secA}, several approximations apply: 
(i) $h_{o,s}\sim -{\bf p}^2/ m \sim H \sim \lQ$ in the semi-hard regions,
(ii) whereas $h_{o,s} \ll H \sim \lQ$ in the potential
regions. Moreover, we can expand the incoming (outcoming) three-momenta 
with respect to the semi-hard ones in $ V_{s,o}^{sh,p}$ and $ V_{s,o}^{p,sh}$. 

By summing up all the contributions, we obtain the same
result as in Sec. \ref{secA}: 
\be 
{\rm Re}\, 
\delta V^{1/\sqrt{m}} = (2C_f+C_A)^2 {4 \over
3\Gamma(9/2)} \pi \, \als^2 \, {\cal E}_{7/2}^E\; {\delta^3({\bf r})
\over m^{3/2}} \,.  
\ee
This is not a coincidence. Note first that the diagram in
Fig. \ref{figen2}d is identical to the one in Fig. \ref{figen}d. The
remaining diagrams in Fig. \ref{figen2} also have a mapping to the
corresponding ones of Fig. \ref{figen}, if we substitute the square
box in the former by a round box linked to an open circle through an octet
propagator. This mapping can be made rigorous from the following equality 
(where $\{|n\rangle^{(0)}\}$ is the gluonic term of a complete set of eigenstates of the static 
NRQCD Hamiltonian, and $E_n^{(0)}$ the corresponding eigenvalues \cite{m,sw}):
\bea
&& \int d^3R\;\; 
{}^{(0)}\langle \underline{0}; {\bf x}_1, {\bf x}_2| \, 
{\bf J}({\bf R}) = {}^{(0)}\langle 0| \, {\bf D}_{{\bf x}_1} 
\, \delta^3 ({\bf x}_1-{\bf x}_2) 
\nn\\
&& \qquad\qquad 
= \sum_{n\neq 0} 
{}^{(0)}\langle 0| \, {\bf D}_{{\bf x}_1} 
\, \delta^3 ({\bf x}_1-{\bf x}_2) |n\rangle^{(0)}\;
{}^{(0)}\langle n| + \dots 
\nn\\
&& \qquad\qquad = \delta^3 ({\bf x}_1-{\bf x}_2) {1 \over \sqrt{N_c}} \sum_{n\neq 0} 
\langle {\rm vac} | \, g{\bf E}({\bf x}_1) 
{1 \over E_0^{(0)} - H}|n\rangle^{(0)} \; {}^{(0)}\langle n|
+ \dots 
\,,
\label{tech}
\eea
where in the last line we have also made use of the fact that 
in the limit ${\bf x}_1 - {\bf x}_2  \to 0$ we have 
$|0\rangle^{(0)} \to 1\!\!{\rm l}_c |{\rm vac}\rangle/\sqrt{N_c}$.
The neglected terms, generically denoted with dots, do not give 
delta-type contributions to the potentials.
From Eq. (\ref{tech}) it follows that the calculation of the diagrams of
Fig. \ref{figen2}a, Fig. \ref{figen2}b and  Fig. \ref{figen2}c
reduces to that one of the diagrams of Fig. \ref{figen}a, 
Fig. \ref{figen}b and  Fig. \ref{figen}c respectively. 
Similarly, for the imaginary part of $\delta V^{1/\sqrt{m}}$ 
the relevant diagrams reduce to those  calculated in Sec. \ref{secdec} 
and shown in Fig. \ref{figde}. It reads:
 \be
{\rm Im} \, \delta V^{1/\sqrt{m}}
=
(2C_f+C_A) {4 \over 3\Gamma(7/2)} \, K_s\, \als\,
 {\cal E}^E_{5/2}\; {\delta^3({\bf r})\over m^{5/2}} \,.
\ee

Here, as well, an analysis for the $P$-wave decays could be done. We
can easily estimate that the leading effects would be at least
$O(m\als/\sqrt{m\lQ})$ suppressed with respect to the contributions
computed in \cite{pw}.

\section{Conclusions}
For heavy quarkonium systems in the strong-coupling regime 
($\lQ \gg mv^2$), the corrections to the static QCD potential
in the Schr\"odinger equation have so far been calculated within a
$1/m$ expansion. We have shown here in a quantitative manner that they are not 
the only contributions to the full potential and have computed the leading non-analytical  corrections in $1/m$. 

Our findings can be summarized in the following corrections to the energy
levels and the $S$-wave matrix elements and decay widths
(the symbols $V$ and $P$ stand for the vector 
and pseudoscalar $S$-wave heavy quarkonium respectively, $n$ is the principal 
quantum number):
\be
\delta E = (2C_f+C_A)^2 {1 \over
3\Gamma(9/2)} \, \als^2 \, {\cal E}_{7/2}^E\;
{|R_{nl}({\bf 0})|^2 \over m^{3/2}}\delta_{l0} 
\,,
\ee
\be
\langle V_Q(nS)|O_1(^3S_1)|V_Q(nS)\rangle=
C_A {|R^V_{n0}({0})|^2 \over 2\pi}
\left(1+  {4 (2C_f+C_A)\over 3\Gamma(7/2)} \, \; {\als\,
 {\cal E}^E_{5/2}\over m^{1/2}}+ O\left({1 \over m}\right)\right)
\,,
\ee
\be
\langle P_Q(nS)|O_1(^1S_0)|P_Q(nS)\rangle=
C_A {|R^P_{n0}({0})|^2 \over 2\pi}
\left(1+  {4 (2C_f+C_A)\over 3\Gamma(7/2)} \, \; {\als\,
 {\cal E}^E_{5/2}\over m^{1/2}}+ O\left({1 \over m}\right)\right)
\,,
\ee
\bea
\Gamma(V_Q (nS) \rightarrow LH) &=&{C_A \over \pi}{|R^V_{n0}({0})|^2 \over m^2}
\nn\\
&&\times
\left[
{\rm Im\,}f_1(^3 S_1)\left(1+{4 (2C_f+C_A)\over 3\Gamma(7/2)} \, \; {\als\,
 {\cal E}^E_{5/2}\over m^{1/2}}+ O\left({1 \over m}\right)\right)
+\cdots
\right], \label{hadrV}
\\
&&
\nn 
\\
\Gamma(P_Q (nS) \rightarrow LH) &=& {C_A \over \pi}{|R^P_{n0}({0})|^2 \over m^2}
\nn\\
&&\times
\left[
{\rm Im\,}f_1(^1 S_0)\left(1+{4 (2C_f+C_A)\over 3\Gamma(7/2)} \, \; {\als\,
 {\cal E}^E_{5/2}\over m^{1/2}}+ O\left({1 \over m}\right)\right)
+\cdots
\right],
\eea
\bea
\Gamma(V_Q (nS) \rightarrow e^+e^-) &=& {C_A \over \pi}{|R^V_{n0}({0})|^2 \over m^2}
\nn\\
&&\times
\left[
{\rm Im\,}f_{ee}(^3 S_1)\left(1+{4 (2C_f+C_A)\over 3\Gamma(7/2)} \, \; {\als\,
 {\cal E}^E_{5/2}\over m^{1/2}}+ O\left({1 \over m}\right)\right) 
+\cdots
\right],
\\
&&
\nn
\\
\Gamma(P_Q (nS) \rightarrow \gamma\gamma) &=&
{C_A \over \pi}{|R^P_{n0}({0})|^2 \over m^2}
\nn\\
&&\times
\left[
{\rm Im\,}f_{\gamma\gamma}(^1 S_0)\left(1+{4 (2C_f+C_A)\over 3\Gamma(7/2)} \, \; {\als\,
 {\cal E}^E_{5/2}\over m^{1/2}}+ O\left({1 \over m}\right)\right) 
+\cdots
\right],
\eea
where $O(1/m)$ stands for corrections (which may be of the same size) that can be
computed within the $1/m$ expansion (see \cite{sw}) and for higher-order corrections.

Let us comment on the size of the new corrections. For the spectrum they are always 
smaller than $mv^3$ and therefore subleading with respect those calculated in \cite{m}. 
For the $S$-wave decay widths their relative size with respect the corrections computed in 
\cite{sw} depends on the size of $\als(\sqrt{m\lQ})$. Under some circumstances, for 
instance $\als \sim v$, the contributions calculated here are the
dominant ones. In any case, the above results fulfil the same factorization properties as
those obtained in \cite{sw}. As a consequence, equations like those
given in Sec. VII of Ref. \cite{sw} still hold. Let us also note that the same non-perturbative
correlator appears in both electromagnetic and hadronic decays.

In this paper we have assumed that the scale $\sqrt{m\lQ}$ 
is much larger than $m\als$. Otherwise we are not allowed to treat the Coulomb potential 
as a perturbation at that scale. This may not be the case for the $\Upsilon$ system 
where one seems to be in the situation $\lQ \sim m\als^2$, which implies 
 $\sqrt{m\lQ} \sim m\als$. In this case, one should
integrate out the three-momentum scale $m\als$ at the same time as the scale
$\sqrt{m\lQ}$. The calculations presented here should be modified by using 
the full Coulomb propagators instead of the free ones in the semi-hard regions.
In addition extra contributions may arise, which are only due to the three-momentum 
scale $m\als$. We do not deal with this issue in this paper, which, however, deserves further studies.

{\bf Acknowledgements.}
We acknowledge support from a CICYT-INFN 2003 collaboration contract.
N.B. acknowledges the support of the Alexander von Humboldt foundation.
A.P. and J.S. are supported by MCyT and Feder (Spain), FPA2001-3598,
by CIRIT (Catalonia), 2001SGR-00065 and by the EU network EURIDICE,
HPRN-CT2002-00311.  A.V. was supported during this work by a Marie-Curie fellowship, 
contract No. HPMF-CT-2000-00733.

\end{document}